\documentclass[a4paper,11pt]{article}
\usepackage{pos}

\usepackage{dcolumn}

\newcommand{\be}{\begin{equation}}
\newcommand{\ee}{\end{equation}}
\newcommand{\bea}{\begin{eqnarray}}
\newcommand{\eea}{\end{eqnarray}}

\newcommand{\bmp}[1]{\begin{minipage}{#1cm}}
\newcommand{\emp}{\end{minipage}}

\newcommand{\bra}{\langle}
\newcommand{\ket}{\rangle}

\newcommand{\Tr}{{\mathrm{Tr}\,}}

\newcommand{\half}{\frac{1}{2}}

\newcommand{\bean}{\begin{eqnarray*}}
\newcommand{\eean}{\end{eqnarray*}}

\title{Interpreting machine learning functions as physical observables}

\ShortTitle{Interpreting machine learning functions as physical observables}

\author*[a,b]{Gert Aarts}
\author[c]{Dimitrios Bachtis}
\author[c,d]{Biagio Lucini}

\affiliation[a]{
Department of Physics, Swansea University, Swansea SA2 8PP, United Kingdom
}
\affiliation[b]{
    European Centre for Theoretical Studies in Nuclear Physics and Related Areas (ECT*) 
   \& \\ Fondazione Bruno Kessler, Strada delle Tabarelle 286, 38123 Villazzano (TN), Italy
}

\affiliation[c]{
Department of Mathematics, Swansea University, Bay Campus, Swansea SA1 8EN, United Kingdom
}

\affiliation[d]{
Swansea Academy of Advanced Computing, Swansea University, Bay Campus, Swansea SA1 8EN, United Kingdom
}

\emailAdd{\{g.aarts, dimitrios.bachtis, b.lucini\}@swansea.ac.uk}

\abstract{
We propose to interpret machine learning functions as physical observables, opening up the possibility to apply ``standard'' statistical-mechanical methods to outputs from neural networks. This includes histogram reweighting and finite-size scaling, to analyse phase transitions quantitatively. In addition we incorporate predictive functions as conjugate variables coupled to an external field within the Hamiltonian of a system, allowing to induce order-disorder phase transitions in a novel manner. A noteworthy feature of this approach is that no knowledge of the symmetries in the Hamiltonian is required.
}

\FullConference{
 The 38th International Symposium on Lattice Field Theory, LATTICE2021
  26th-30th July, 2021
  Zoom/Gather@Massachusetts Institute of Technology
  }

\begin{document}

\maketitle

\section{Introduction}

The application of machine learning (ML) in the physical sciences has seen a dramatic increase in recent years, see e.g.\ Ref.\ \cite{Carleo:2019ptp} for a broad overview. In this contribution, we will discuss some of the work done in Swansea in the past year \cite{Bachtis:2020dmf,Bachtis:2020ajb,Bachtis:2020fly}; in a parallel talk Ref.\ \cite{Bachtis:2021xoh} was presented, and hopefully at next year's Lattice conference Ref.\ \cite{Bachtis:2021eww} can be discussed in person.

This contribution has three take-home messages, namely
\begin{enumerate}
\item one can (should) interpret ML outputs as physical observables;
\item critical behaviour  can be derived from neural network observables only;
\item new Hamiltonians can be constructed with the inclusion of ML predictive functions.
\end{enumerate}
We will discuss these three messages and their implications in the following sections.

\section{ML outputs as physical observables}

\begin{figure}[b]
\begin{center}
  \includegraphics[width=0.9\textwidth]{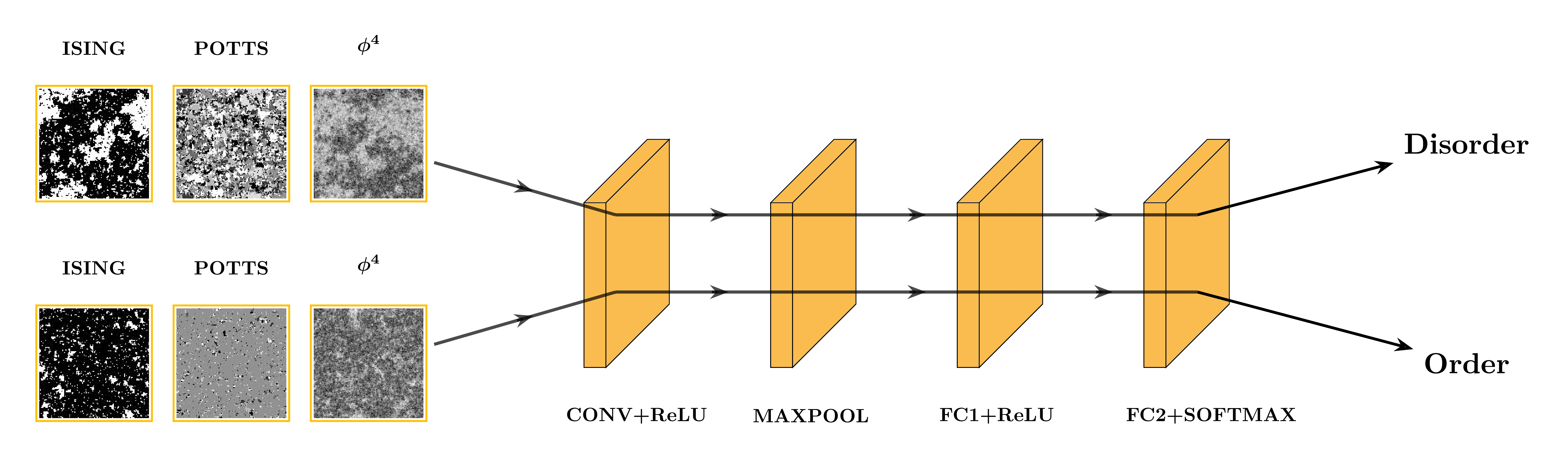}
\caption{Example of a convolutional neural network (CNN) used for phase classification. 
\label{fig:cnn}
}
\end{center}
\end{figure}

As is well known, ML excels in pattern finding. In the context of statistical physics and phases of matter, this has been demonstrated in the by-now classic paper \cite{phasesofmatter}, in which it was demonstrated that ML can classify phases and phase transitions in the two-dimensional Ising model and other lattice systems. The task is formulated as a supervised learning problem: one generates (or otherwise obtains) configurations of spins on a two-dimensional lattice for a variety of temperatures or couplings. Configurations deep in the ordered and the disordered phases are labelled accordingly and are used to train the ML algorithm, i.e.\ adjusting the parameters in the neural network of choice such that the desired classification is correctly reproduced for the training set. New, unseen configurations at intermediate temperatures or couplings are then provided to the network, which returns the probability to be in the (dis)ordered phase. Repeating this on lattices of various sizes allows one to study critical scaling and determine properties of the transition in the infinite-volume limit. This approach has been used in the Ising model, $q$-state Potts models with varying $q$, scalar $\phi^4$ theory, etc. An example of a convolutional neural network used in our work is shown in Fig.\ \ref{fig:cnn}. 

Following Ref.\ \cite{phasesofmatter}, the approach has been repeated and extended many times and is by now a well-established procedure. The obvious question is therefore what can be added in this direction.
As Fig.\ \ref{fig:cnn} illustrates, the input in the network are configurations, which have been generated according to the Boltzmann weight $e^{-\beta H}/Z$, with $H$ the Hamiltonian of the system, $\beta$ the inverse temperature or coupling, and $Z$ the partition function. By considering configurations at fixed $\beta$, the process of computing the outcome of the neural network is identical to how observables are computed in numerical simulations in statistical physics, namely by averaging over configurations. However, the observable in this case is not expressed directly in terms of the degrees of freedom, such as e.g.\ the magnetisation is, but is instead a rather complicated quantity, constructed as described above.
Since the outcome is the probability of being in the (dis)ordered phase, it nevertheless still has the potential to act similar to an  ``order parameter'' in the statistical system.

The reasoning above leads to the proposition that the output of a neural network should be interpreted as an observable in a statistical system  \cite{Bachtis:2020dmf}. This observation is of interest, since it opens up the possibility to use ``standard'' numerical or statistical methods, such as histogram reweighting.
Let $P_i$ denote the neural network prediction a configuration with energy $E_i$ is in the broken phase, with $0\leq P_i\leq 1$.
Averaging over configurations at fixed $\beta$ then yields the expectation value as provided by the neural network,
\be
\bra P\ket = \frac{1}{Z} \sum_i P_i e^{-\beta E_i},
\ee
since the configurations labeled by $i$ are Boltzmann distributed. One can now use standard histogram reweighting \cite{FS1} to extrapolate from a given $\beta_0$ to other $\beta$ values,  according to 
\be
\bra P\ket(\beta) = \frac{ \sum_i P_i e^{-(\beta-\beta_0) E_i}}{\sum_i e^{-(\beta-\beta_0) E_i}},
\ee
and obtain continuous dependence on the coupling constant.

\begin{figure}[h]
\begin{center}
  \includegraphics[width=0.8\textwidth]{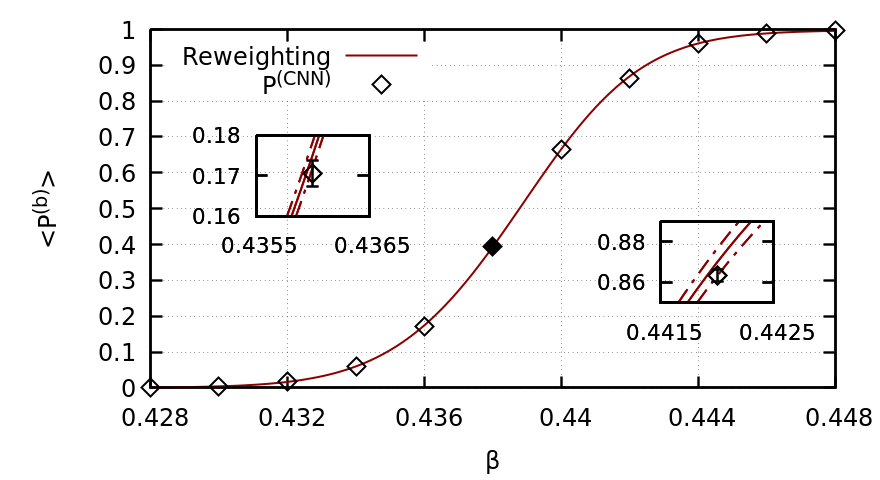}
\caption{Probability $\bra P\ket$ to be in the broken (ordered) phase, as a function of the coupling $\beta$ in the two-dimensional Ising model on a $V=128^2$ lattice. The open and filled diamonds are predictions from the neural network, while the line is obtained via histogram  reweighting, starting from the filled diamond at $\beta_0=0.438$.
\label{fig:Ising}
}
\end{center}
\end{figure}

This approach is illustrated in Fig.\ \ref{fig:Ising}. The CNN shown in Fig.\ \ref{fig:cnn} is trained deeply in the (dis)ordered phases, at    $\beta \leq 0.41$ and $\beta \geq 0.47$. Predictions are made at intermediate values of $\beta$, indicated with the diamonds. The line is obtained employing histogram reweighting, using only the datapoint at $\beta_0 =0.438$ (which is also obtained from the CNN). It is seen that the line is in agreement with the direct predictions from the CNN and interpolates smoothly between the disordered and ordered phases. 
The insets indicate agreement within statistical uncertainty between the two approaches. More details can be found in Ref.\ \cite{Bachtis:2020dmf}.

\begin{figure}[t]
\begin{center}
  \includegraphics[width=0.49\textwidth]{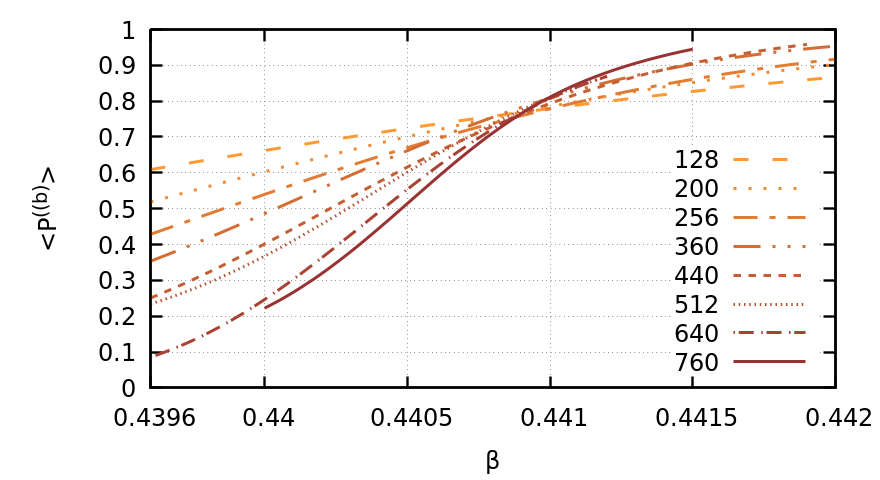}
  \includegraphics[width=0.49\textwidth]{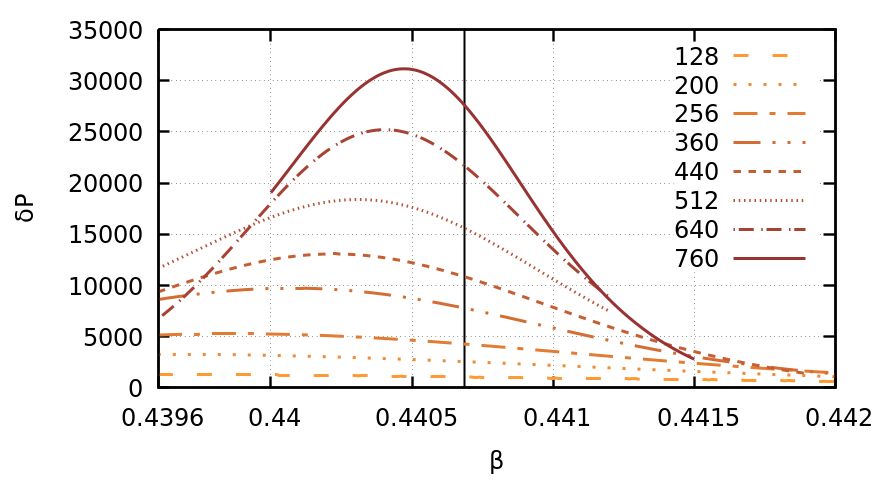}
\caption{Finite-size scaling using histogram reweighing in the Ising model, for the probability to be in the ordered phase $\bra P\ket$ (left) and its susceptibility $\delta P = \beta V \left(\bra P^2\ket - \bra P\ket^2 \right)$ (right). 
\label{fig:Ising2}
}
\end{center}
\end{figure}

All results in Fig.\ \ref{fig:Ising} are predictions from the CNN, with or without reweighting. In order to see whether they give a quantitative prediction for properties of the transition, we carry out a finite-size scaling analysis using the results obtained with histogram reweighting. The results are shown in Fig.\ \ref{fig:Ising2} for the probability $\bra P\ket$ (left) and its susceptibility $\delta P = \beta V \left(\bra P^2\ket - \bra P\ket^2 \right)$ (right). We will use these data to analyse critical scaling in the next section.

\section{Critical behaviour from neural network observables only}

\begin{figure}[h]
\begin{center}
  \includegraphics[height=0.24\textwidth]{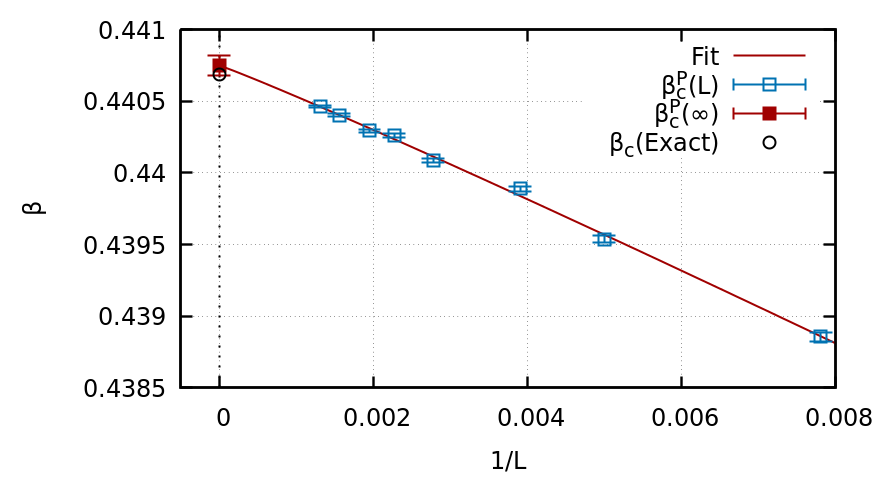}
  \includegraphics[height=0.24\textwidth]{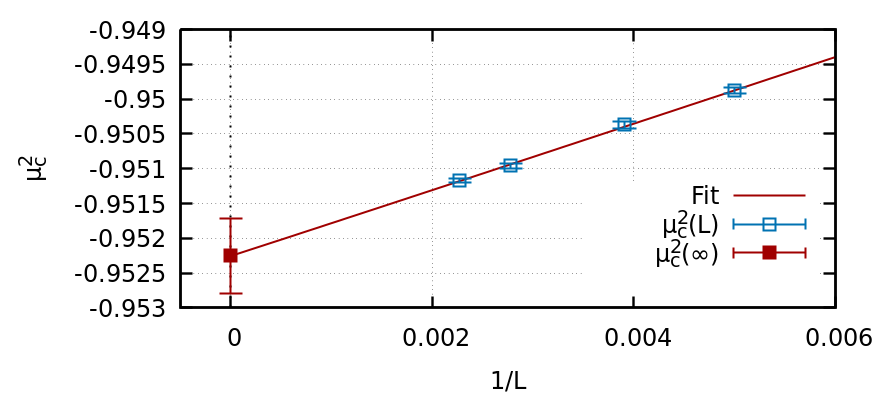}
\caption{Determination of the critical parameters $\beta_c$ and $\mu^2_c$ in the two-dimensional Ising model (left) and the $\lambda\phi^4$ theory (right) respectively  using finite-size scaling of CNN predicted values.
\label{fig:extrapolation}
}
\end{center}
\end{figure}

Finite-size scaling should give us access to the critical coupling $\beta_c$ and critical exponents in the infinite-volume limit. Recall that in the two-dimensional Ising model, $\beta_c = \half \ln(1+ \sqrt{2})$, while the correlation length and susceptibility diverge as 
 $\xi\sim |t|^{-\nu}$ and $\chi\sim  |t|^{-\gamma}$, where $t$ is the reduced coupling, $t=(\beta_c-\beta)/\beta_c$. In a finite system of linear size $L$, this implies 
\be
|t| \sim  \xi^{-1/\nu} \sim L^{-1/\nu}, \qquad\qquad \chi\sim  |t|^{-\gamma}\sim L^{-\gamma/\nu}.
\ee
We use these relations to extract the critical quantities from the observables determined within the neural network, as shown in Fig.\ \ref{fig:Ising2}. Note in particular that the susceptibility is not the magnetic susceptibility, but rather the one linked to the network probability $\bra P\ket$. The result for the volume dependence of $\beta_c$ is shown in Fig.\ \ref{fig:extrapolation} (left) and the obtained values for $\beta_c, \nu$ and $\gamma/\nu$ are given in Table \ref{tab:table2}. We observe good agreement with the expected results within the statistical uncertainty. 

In order to demonstrate that this approach also works in a theory with continuous degrees of freedom, we repeat the exercise in the two-dimensional $\lambda\phi^4$ scalar field theory \cite{Bachtis:2020ajb}. In this case, exact results for the critical couplings are not available, but it is expected that the model is in the same universality class as the Ising model. We fixed the coupling $\lambda=0.7$ and vary the mass parameter $\mu^2$ to locate the transition. Again this is repeated on various volumes. The result for the volume dependence of $\mu^2_c$ is shown in Fig.\ \ref{fig:extrapolation} (right), and the obtained values for $\mu^2_c, \nu$ and $\gamma/\nu$ are given in Table \ref{tab:table2}. The value of the critical mass parameter agrees with the one found in the literature \cite{Schaich:2009jk}, while the value of the critical exponents confirm the universality class.

\begin{table}[t]
\begin{center}
\begin{tabular}{| c | ccc | }
\hline
 Ising model 		& $\beta_{c}$ 						& $\nu$ 	& $\gamma/\nu$ \\
\hline
CNN+Reweighting 	& 0.440749(68) 					 & 0.95(9) & 1.78(4) \\
Exact 			& $\half \ln(1+\sqrt{2})  \approx 0.440687$ & $1$ 	& $7/4  =1.75$  \\
\hline
$\phi^4$ theory 	& $\mu_{c}^{2}$ 					 &$\nu$ 	& $\gamma/\nu$ \\
\hline
CNN+Reweighting 	&  -0.95225(54)  					 & 0.99(34) & 1.78(7) \\
\hline
\end{tabular}
\end{center}
\caption{\label{tab:table2}
Numerical estimates for the critical parameters $\beta_c$  in the Ising model and $\mu_{c}^{2}$ in the $\lambda\phi^4$ theory at fixed $\lambda=0.7$, as well as for the critical exponents $\nu$ and $\gamma/\nu$, in the infinite-volume limit.
}
\end{table}

We conclude that critical behaviour can be derived from observables obtained  within the neural network formulation, with adequate precision.

\section{Extend Hamiltonians with ML predictive functions}

We can now take these insights one step further and prepare new Hamiltonians, by including the ML predictive function in the Hamiltonian itself \cite{Bachtis:2020fly}, i.e.\ we couple the ML predictive function to a conjugate source $Y$ which, when varied to positive and negative values, can drive the system into the ordered/disordered phases -- in both directions -- and also allows us to extract an additional critical exponent.

Let us denote the ML predictive function, i.e\ the probability to be in the ordered phase, with $f$, where $0 \leq f\leq 1$. Since it is an intensive quantity, it is multiplied with the volume $V$ and added to the system Hamiltonian $H$, coupled to the conjugate source $Y$,  to arrive at $H_Y =H-VfY$.
This formulation can be used to define a generating function, 
\be
\bra f\ket = \frac{1}{\beta V}\frac{\partial \ln Z_Y}{\partial Y},
\qquad\quad
Z_Y = \Tr e^{-\beta H_Y},
\qquad\quad
H_Y =H-VfY,
\ee
etc, but more importantly, it can induce symmetry breaking without explicit symmetry breaking.
 
\begin{figure}[h]
\begin{center}
  \includegraphics[width=0.49\textwidth]{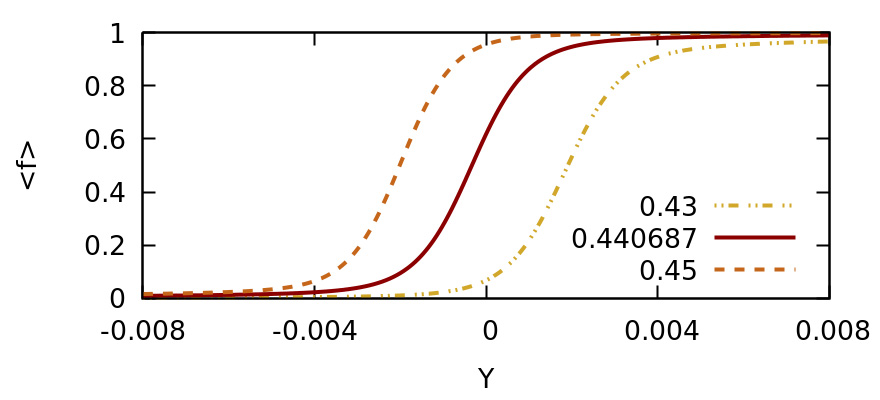}
  \includegraphics[width=0.49\textwidth]{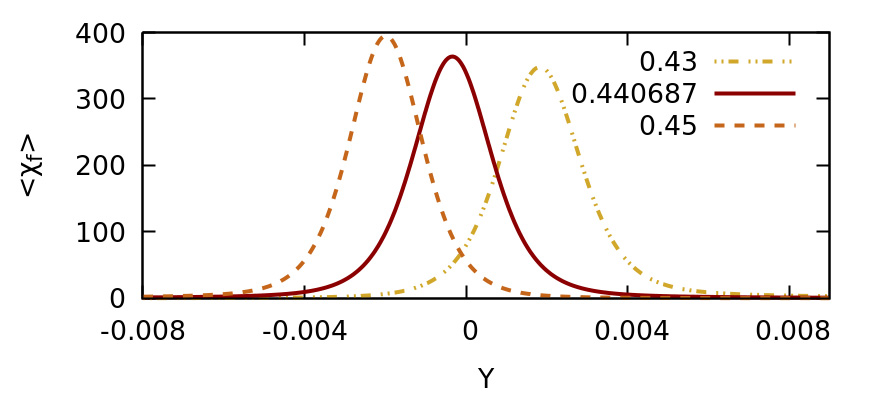}
\caption{Response to the external field $Y$ of the probability to be in the ordered phase $\bra f\ket$ (left) and its susceptibility $\chi_f$ (right) in the two-dimensional Ising model on a $64^2$ lattice, for three values of the coupling~$\beta$.
}
\label{fig:external}
\end{center}
\end{figure}

This is demonstrated in Fig.\ \ref{fig:external}, where the response of $\bra f\ket$ and its susceptibility $\chi_f$ are shown as $Y$ is varied. The three values of $\beta$ correspond to the symmetric phase ($\beta=0.43$), the known critical coupling ($\beta_c=0.440687$), and the broken-symmetry phase ($\beta=0.45$) at $Y=0$. We note that the results for nonzero $Y$ are again obtained using histogram reweighting.
We indeed observe that as $Y$ is varied, the system moves seamlessly from one phase into the other, and in both directions as the sign of $Y$ is changed. This should be contrasted with the inclusion of an external magnetic field, in which the symmetry is always broken explicitly, giving preference to the symmetry-broken phase. 

To bring this discussion to an end, we finally note that an external field introduces a new critical exponent. In the case of an external magnetic field, the correlation length varies as $\xi \sim  | h_{\rm magnetic} |^{-\theta}$, where in the Ising model $\theta = 8/15$. Since it is not guaranteed that the dependence on $Y$ will be captured by the same exponent, we denote it with $\theta_Y$ and conjecture that $\xi \sim  | Y|^{-\theta_Y}$. We have investigated this scaling numerically, using blocking and a renormalisation group (RG)  analysis \cite{Bachtis:2020fly}. This study gives access to $\beta_c, \nu$ and $\theta_Y$, with the results shown in Table \ref{tab:table3}. We observe excellent agreement and confirm that $\theta_Y$ can indeed be identified with the exponent $\theta$.

We reiterate that this analysis only used quantities derived from the neural network, that there is no need for knowledge of the order parameter, the pattern of symmetry breaking, the physical meaning of external field, or the introduction of explicit symmetry breaking.

\begin{table}[h]
\begin{center}
\begin{tabular}{| c | ccc | }
\hline
  Ising model		& $\beta_{c}$ & $\nu$ & $\theta_Y, \theta$ \\
\hline
NN$+$RG 	& 0.44063(21) 		& 1.02(2) 	& $\theta_Y = 0.534(3)$\\
Exact 		& $\half \ln(1+\sqrt{2}) \approx 0.440687$ 	& $1$ 	& $\theta=8/15 \approx 0.5333$ \\
\hline
\end{tabular}
\caption{Numerical estimates for $\beta_c$ and the critical exponents $\nu$ and $\theta_Y$ in the Ising model, using the Hamiltonian extended  with the ML predictive function coupled to a source. 
}
\label{tab:table3}
\end{center}
\end{table}

\section{Summary}

In this contribution we reviewed our proposal to identify outputs from neural networks as observables in statistical physics. The first consequence is that this allowed us to introduce histogram reweighting to be employed in supervised machine learning. This yields the possibility to extract critical properties from a finite-size scaling analysis using quantities derived from neural networks alone, to relatively high precision. Detailed knowledge of symmetries and order parameters is not required.
By coupling neural network outputs to an external source, we can drive the system into the ordered/disordered phases,  unlike in the standard way of using an external magnetic field which always breaks the symmetry explicitly.

We hope these observations can lead to new quantitative studies of phase transitions based on a synergistic relation between machine learning and statistical mechanics,  with ideas and inspiration which is of interest to both communities.

 
 \vspace*{0.5cm}
 
\noindent
{\bf Acknowledgments} -- 
The authors received funding from the European Research Council (ERC) under the European Union's Horizon 2020
research and innovation program under Grant Agreement No.\ 813942. GA and BL are supported in part by the UKRI Science and Technology Facilities Council (STFC) Consolidated Grant No.\ ST/P00055X/1 and ST/T000813/1. BL is supported in part by the Royal Society Wolfson Research Merit Award No.\ WM170010 and by the Leverhulme Foundation Research Fellowship No.\ RF-2020-461/9. Numerical simulations have been performed on the Swansea SUNBIRD system. This system is part of the Supercomputing Wales project, which is partly funded by the European Regional Development Fund (ERDF) via the Welsh Government.


\end{document}